\documentclass[preprint,showpacs,amsmath,amssymb]{revtex4}

\usepackage{graphicx}
\usepackage{dcolumn}
\usepackage{bm}

\begin{document}

\title{Low Temperature Thermodynamic Properties of the Heavy Fermion Compound YbAgGe Close to the Field-Induced Quantum Critical Point}

\author{Y. Tokiwa$^\P$, A. Pikul$^{\P\ddag}$, P. Gegenwart$^\P$, F. Steglich$^\P$, S.L. Bud'ko$^\ast$, and P.C. Canfield$^\ast$}
\affiliation{$^\P$Max-Planck Institute for Chemical Physics of Solids, D-01187 Dresden, Germany\\
$^\ddag$Institute of Low Temperature and Structure Research, Polish Academy of Sciences, 50-950 Wroc{\l}aw, Poland\\
$^\ast$Ames Laboratory US DOE and Department of Physics and Astronomy, Iowa State University, Ames, IA 50011, USA}

\date{\today}

\begin{abstract}
We present temperature and field dependent heat capacity and magnetization data taken at temperatures down to 50
mK and in an applied magnetic field up to 11.5 Tesla for YbAgGe, a heavy-fermion compound with a field induced
quantum critical point. These data clearly indicate that the same electronic degrees of freedom are responsible
for the features seen in both specific heat and magnetization data. In addition, they further refine the different
boundaries suggested for the $H - T$ phase diagram of YbAgGe through previous, magneto-transport measurements, and
allow for further understanding of different phases on the $H - T$ phase diagram, in particular, clearly
disconnecting the field-induced quantum critical point in YbAgGe from any sort of saturation of the Yb moment in
higher applied magnetic field.
\end{abstract}


\maketitle

\section{Introduction}

YbAgGe was recently recognized to be one of the few stoichiometric Yb-based heavy fermion compounds that
demonstrates field-induced non-Fermi-liquid (NFL) behavior as evidenced by thermodynamic and transport
measurements in an applied magnetic field \cite{bud04a}. In zero applied field YbAgGe has two (small magnetic
moment) magnetic ordering transitions, at $\approx 1$ K and $\approx 0.65$ K \cite{bud04a,mor04a,ume04a}, with the
latter one being first order. The magnetic ordering temperatures are well separated from the Kondo temperature ($T
_K \sim 25$ K) that is itself well below the estimated first excited CEF level ($T_{CEF} \sim 60-100$ K)
\cite{kat04a,mat04a}. The anisotropic $H - T$ phase diagrams for YbAgGe were constructed based on heat capacity
measurements down to 400 mK and magnetotransport measurements down to temperatures of $\approx 50$ mK
\cite{bud04a,ume04a,bud05a,nik05a,bud05b}, and the nature of some of the magnetic phases has been probed by recent
neutron scattering measurements \cite{fak05a,fak05b}. Whereas the general outline of these phase diagrams is
similar to the generic $H - T$ phase diagram for the materials with a quantum critical point (QCP) (see {\it e.g.}
Ref. \onlinecite{con89a}), (i) they have an additional line, defined by a clear anomaly in the Hall effect
measurements \cite{bud05a,bud05b}, that apparently originates at the QCP (a similar line was also suggested for
YbRh$_2$Si$_2$ \cite{pas04a}); (ii)the apparent non-Fermi-liquid (NFL) region does not collapse to a point in the
$T \to 0$ limit, but spans a magnetic field range of several Tesla \cite{bud04a,nik05a,bud05b}.

With this complex phase diagram in mind and lacking detailed neutron scattering data covering a wide $H - T$
domain, field- and temperature-dependent magnetization measurements gain an additional importance in the case of
the field-induced QCP, allowing for the clear identification of the saturated paramagnetic state (if reached) and
often giving supplementary information on the intermediate, magnetic-field-induced, phases. Whereas in the
previous publications magnetotransport data have been collected for $T \geq 50$ mK and $H \leq 18$ T
\cite{bud04a,ume04a,bud05a,nik05a,bud05b} and $C_p(T,H)$ results so far exist in the $T \geq 400$ mK and $H \leq
14$ T domain \cite{bud04a,ume04a}, systematic magnetization data have only been collected for $T \geq 1.8$ K. To
address this critical gap in the data set, the core of this work is a set of $M(H,T)$ data covering the vicinity
of the QCP. In addition, in order to compare these magnetization data with another, key, thermodynamic quantity,
additional heat capacity data were taken over the same $H - T$ range.

\section{Experimental}

YbAgGe single crystals in the form of clean, hexagonal-cross-section rods of several mm length and up to 1 mm$^2$
cross section were grown from high temperature ternary solutions rich in Ag and Ge (see \cite{mor04a} for details
of the samples' growth). Their structure and full site-occupancy, without any detectable site-disorder, were
confirmed by single crystal X-ray diffraction \cite{moz04a}. DC-magnetization of a single crystalline sample of
YbAgGe was measured down to $\approx 0.05$ K and in high fields up to 11.5 T using a high-resolution capacitive
Faraday magnetometer \cite{sak94a}. The specific heat measurements were carried out down to $\approx 0.05$ K in
magnetic fields up to 11 T using the compensated quasi-adiabatic heat pulse method \cite{wil04a}. For both
measurements the magnetic field was applied perpendicular to the crystallographic $c$-axis, {\it i.e.} along the
basal, $ab$ plane.

\section{Results and Discussion}

\begin{figure}
\begin{center}
\includegraphics[angle=0,width=120mm]{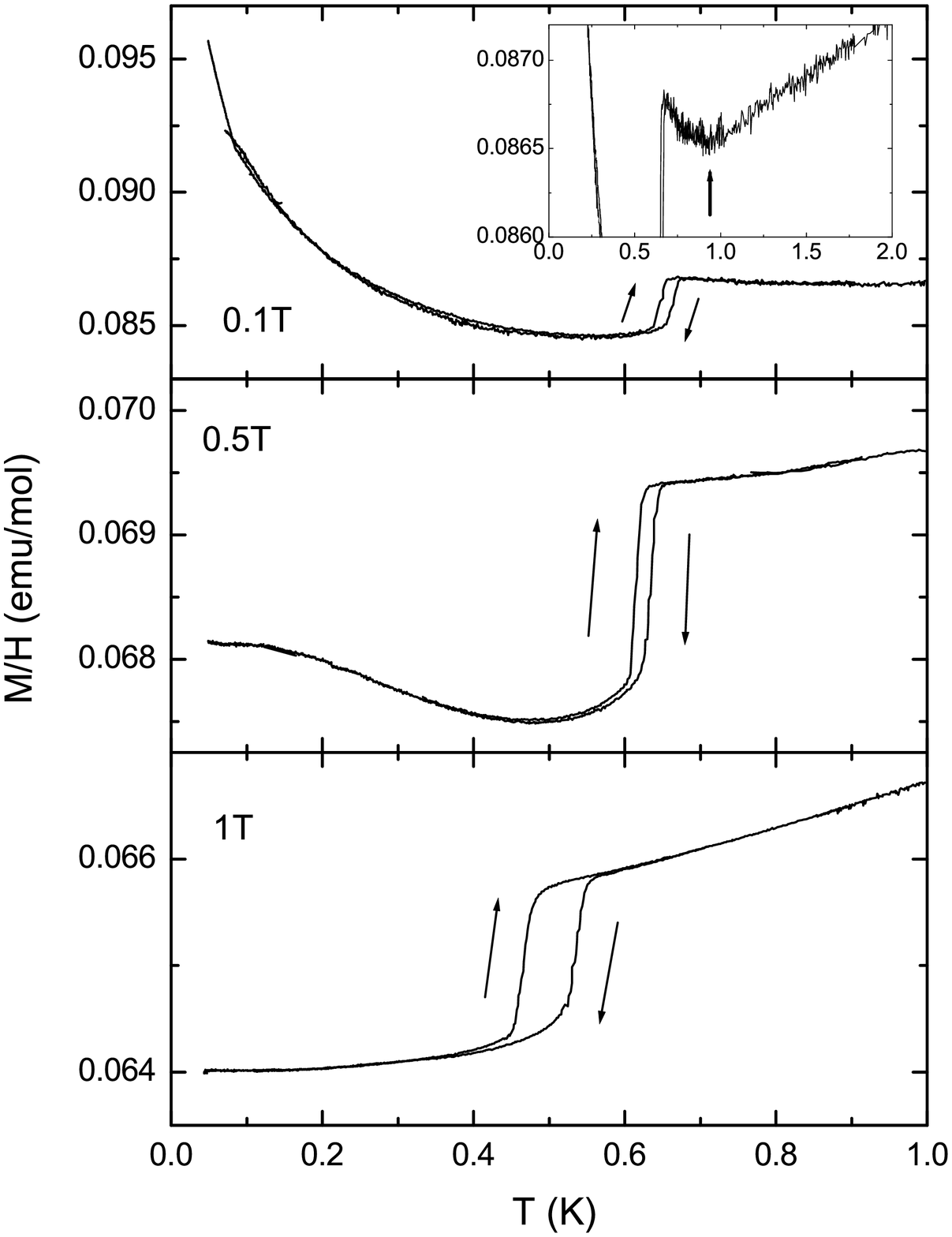}
\end{center}
\caption{Representative plots of the low-field temperature dependent susceptibility of YbAgGe ($H \| ab$, $H \leq
1$ T) below 1 K. Arrows mark data taken on increase/decrease of temperature. Inset shows break of slope near 1 K
in $H = 0.1$ T data.}\label{fi1a}
\end{figure}

\begin{figure}
\begin{center}
\includegraphics[angle=0,width=120mm]{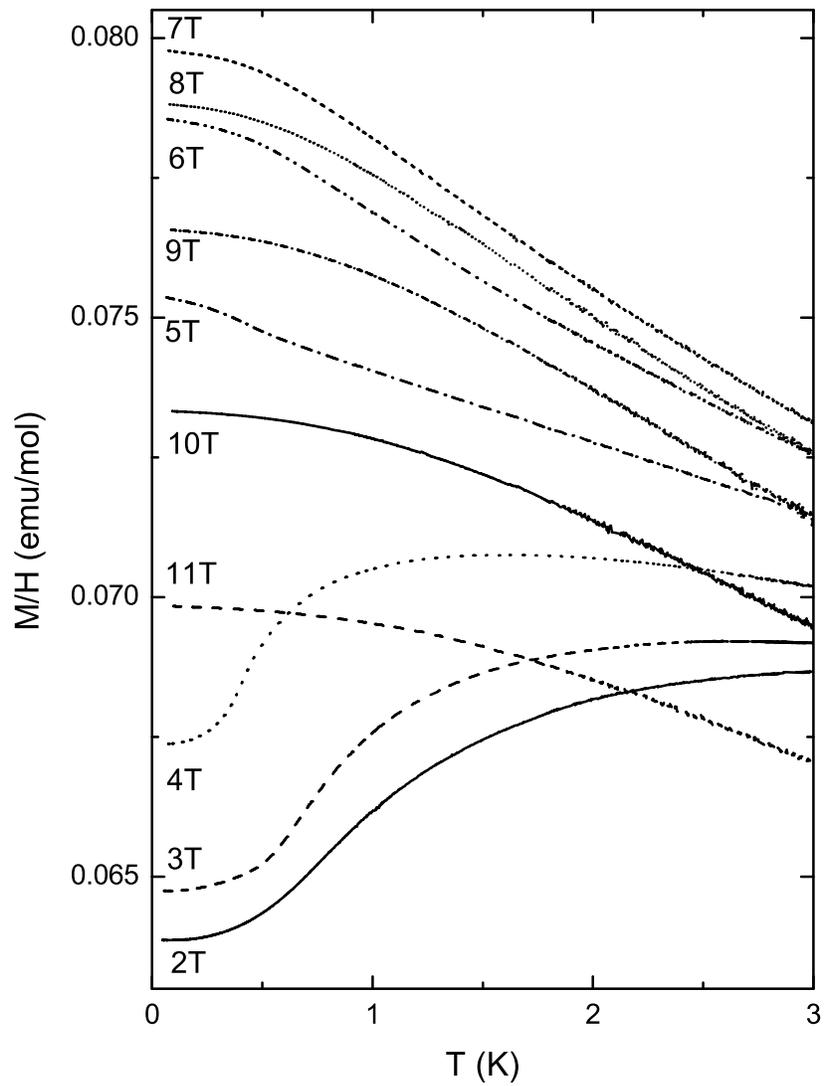}
\end{center}
\caption{Temperature dependent susceptibility of YbAgGe in magnetic fields between 2T and 11 T ($H \| ab$) below 3
K. Data taken on temperature increase.}\label{fi1b}
\end{figure}

A representative set of temperature-dependent $M/H$ data is shown in Figs. \ref{fi1a},\ref{fi1b}. The small, low
temperature upturn in the low-field data (Fig. \ref{fi1a}) is possibly caused by a very small amount of
paramagnetic impurities (consistent with the shape of $M(H)$ in the low-field region - not shown here). This
upturn disappears by $H =1$ T. The lower, first order, magnetic transition (\cite{bud04a,mor04a,ume04a}) is
clearly seen in the low field susceptibility data (Fig. \ref{fi1a}). It shifts to lower temperatures with an
increase of applied field, and the width of the hysteresis increases. It is worth mentioning that for $H = 0.1$ T
a relatively sharp change of slope is observed slightly below 1 K (insert to Fig. \ref{fi1a}). This feature marks
the higher temperature magnetic transition.

The lower temperature transition is not detectable in the temperature-dependent susceptibility measurements taken
in applied fields of 2 T and higher (Fig. \ref{fi1b}). The functional behavior of the susceptibility clearly
changes when applied field increases from 2 T to 11 T. Up to 4 T a magnetic transition can be followed by
monitoring the maximum in $d(M/H)/dT$; For $H \geq 6$ T the magnetic susceptibility increases with decrease of
temperature with the trend to saturation below $\sim 1$ K. The 5 T susceptibility behaves similarly to the higher
field ones with an additional feature near 0.5 K.

All in all the temperature-dependent susceptibility data is consistent with the $H - T$ phase diagram for $ H \|
ab$ suggested in Refs. \onlinecite{bud04a,ume04a,bud05a,nik05a,bud05b} (shown with these data added in Fig.
\ref{fi3} below) with qualitative changes in the functional behavior when the first order transition is suppressed
($H \sim 2$ T) and at the critical field, $H_{QCP}$, when the second order phase transition is suppressed.

\begin{figure}
\begin{center}
\includegraphics[angle=0,width=120mm]{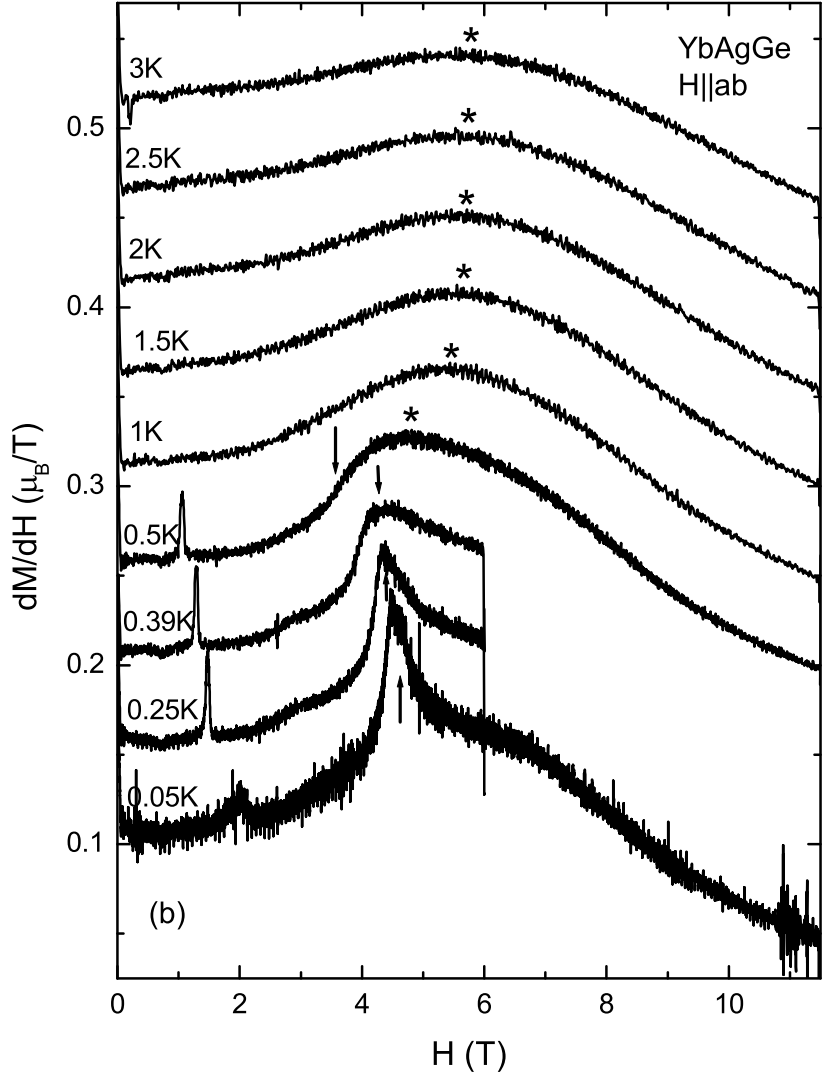}
\end{center}
\caption{(a) Representative plots of low-temperature field-dependent magnetization, $M(H)$, of YbAgGe ($H \| ab$).
Curves are shifted along the $y$-axis by multiplicative factors of 0.1 $\mu_B$ for clarity; (b) $dM/dH$
derivatives for the representative $M(H)$ taken on increase of the magnetic field - curves for $T \geq 0.25$ K
have an offset by a multiplicative of 0.05 $\mu_B$/T for clarity  (see text for the discussion of different
features).}\label{fi2}
\end{figure}

A representative $M(H)$ data set is presented in Fig. \ref{fi2}(a). A well-defined kink at $\sim 4.5$ T is evident
in the raw data for $T = 0.05$ K curve. This anomaly broadens with increasing temperature. The magnetization at
11.5 T, for the whole range of temperatures, shows only the initial signs of possibly approaching a saturation.
This set of data is consistent with the published $T = 0.45$ K \cite{ume04a} as well as the $T = 2$ K
\cite{bud04a} $M(H)$ curves.

\begin{figure}
\begin{center}
\includegraphics[angle=0,width=120mm]{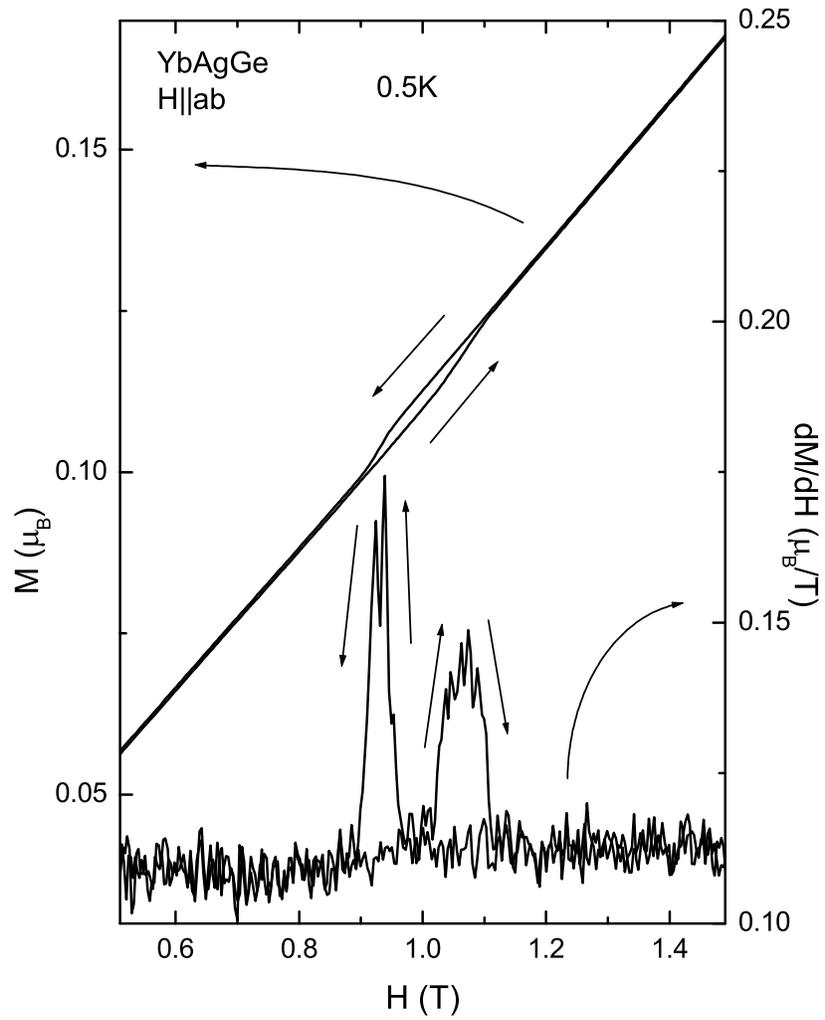}
\end{center}
\caption{Example of hysteretic low field/low temperature transition as seen in raw $M(H)$ data and in the
derivative $dM/dH$.}\label{fihy}
\end{figure}

The low field, first order phase transition can also be clearly detected in the $M(H)$ data. This is shown in
Figure \ref{fihy} where an enlargement of a field-up / field-down sweep is shown. This transition is even more
clearly seen in the $dM/dH$ plot shown on the same panel, the right hand axes.

On a broader scale, the derivatives of the magnetization isotherms, $dM/dH$ (Fig. \ref{fi2}(b)), allow for a
closer look at the features in magnetization and their evolution with temperature. As discussed above, the narrow
peaks in $dM/dH$ at and below 2 T (for $T \leq 0.5$ K) indicate the phase line associated with the lower, first
order, magnetic transition. In addition, an asymmetric peak below 5 T is clearly seen in the 0.05 K, 0.25 K and
0.39 K curves (marked by arrows); this feature evolves into a less pronounced, high slope feature at 0.5 K (marked
with an arrow) with a broad maximum ($\ast$) appearing at $\sim 5$ T. This broad maximum persists up to at least 3
K. Additionally, a smaller maximum ($\triangledown$) can be observed for the $dM/dH$ data taken at 0.25 K and 0.39
K near 3 T. In higher fields, {\it e.g.} $H \approx 7$ T for the 0.05 K curve, a broad shoulder in $dM/dH$ is also
observed. This shoulder is only marginally detectable for $T = 0.5$ K and lack of high field data for the
intermediate temperatures does not allow to follow its evolution with temperature.

\begin{figure}
\begin{center}
\includegraphics[angle=0,width=120mm]{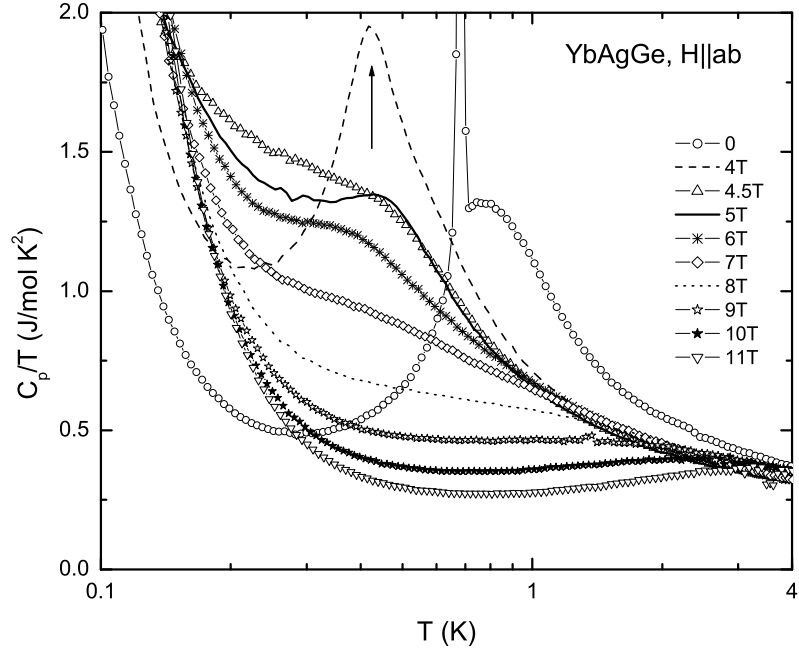}
\end{center}
\caption{Low temperature part of the temperature-dependent heat capacity of YbAgGe ($H \| ab$) plotted as $C_p/T$
{\it vs.} $T$ on a semi-log plot. Arrow marks magnetic transition in $H = 4$ T curve.}\label{fi5}
\end{figure}

The low temperature part of the temperature-dependent heat capacity in constant applied magnetic field is shown in
Fig. \ref{fi5}. All the data below $\sim 200$ mK show an upturn that corresponds to nuclear contributions to the
specific heat. The data above $\sim 400$ mK are consistent with the previously reported heat capacity data
\cite{bud04a}. For $H = 4$ T, the onset of magnetic order can still be clearly seen in the heat capacity data
(marked with an arrow in Fig. \ref{fi5}).

\begin{figure}
\begin{center}
\includegraphics[angle=0,width=120mm]{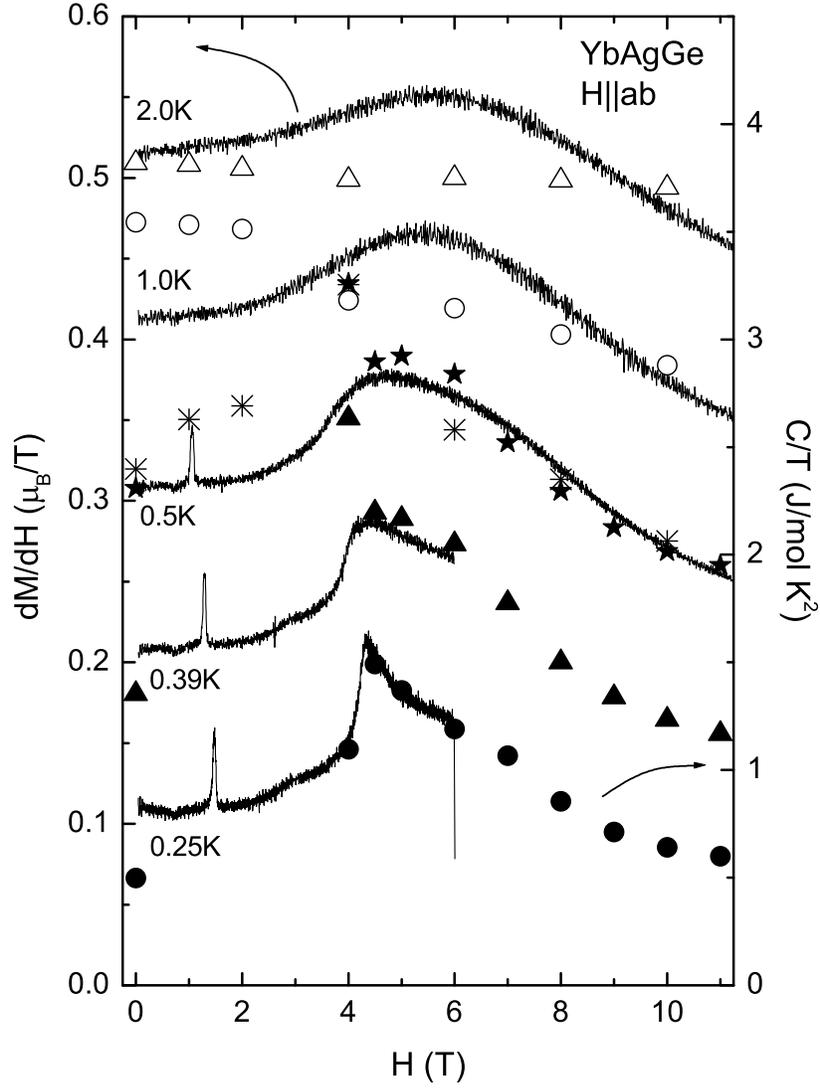}
\end{center}
\caption{Low temperature differential susceptibility, $dM/dH$, and Sommerfeld coefficient, $C/T$, as a function of
applied magnetic field. Data for $T \geq 0.39$ K are shifted vertically by multiplicative of 0.1 $\mu_B$/T
($dM/dH$) and multiplicative of 0.825 J/mol K$^2$ ($C/T$) respectively for clarity. Asterisks and open symbols -
$C/T$ obtained using the data from Ref. \onlinecite{bud04a}, other symbols - this work.}\label{fi4}
\end{figure}

\begin{figure}
\begin{center}
\includegraphics[angle=0,width=80mm]{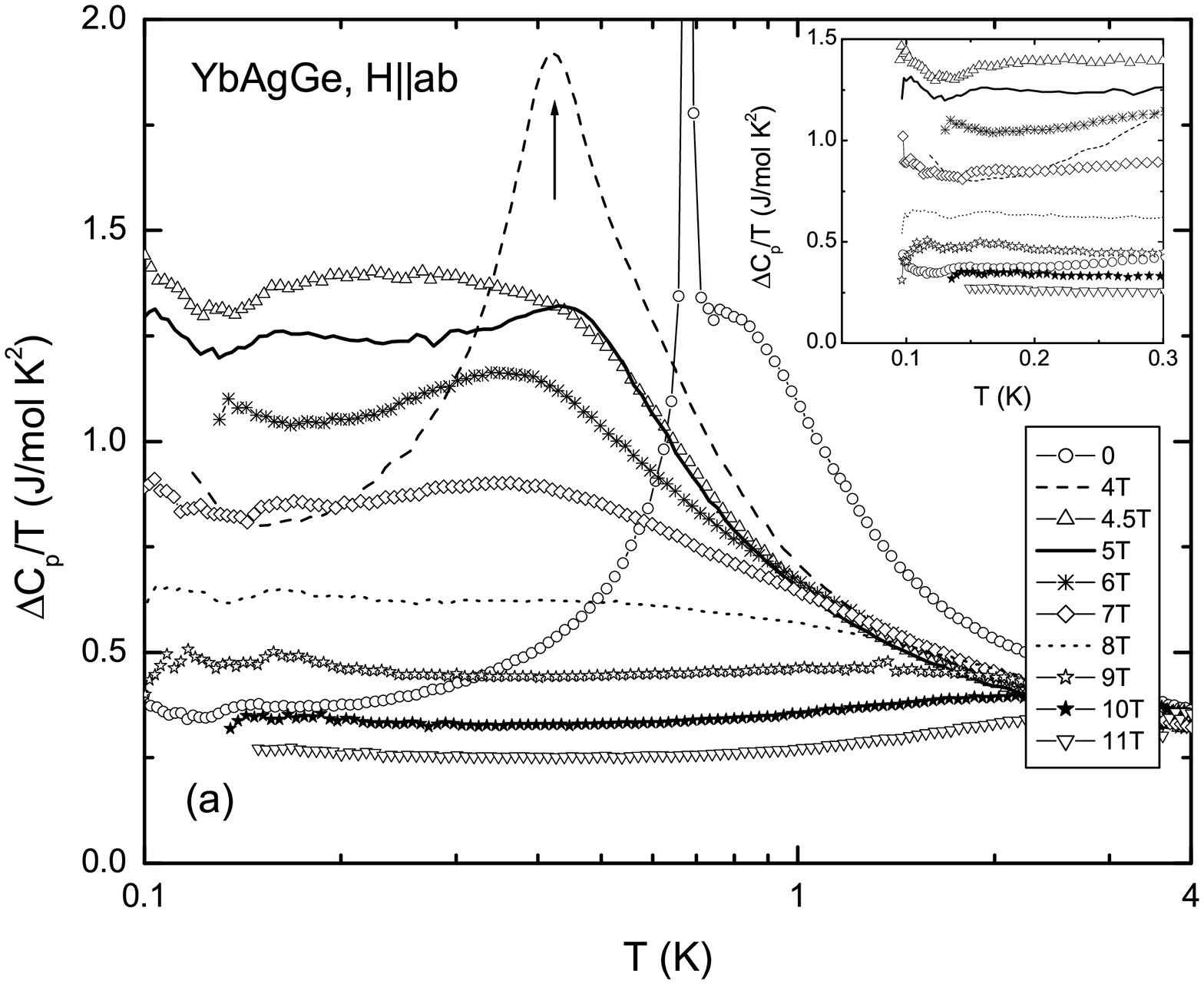}
\includegraphics[angle=0,width=80mm]{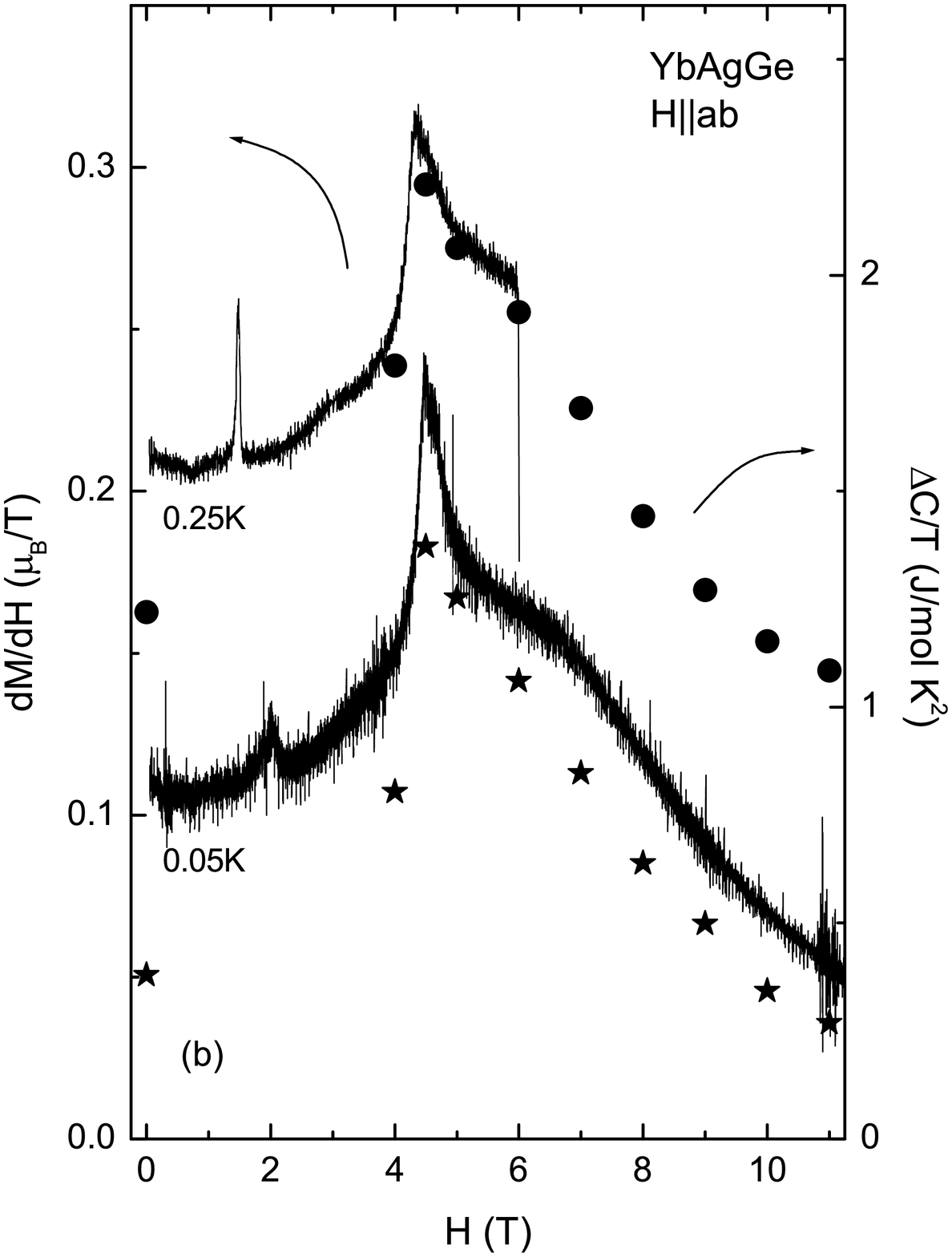}
\end{center}
\caption{(a) Low temperature part of the temperature-dependent heat capacity of YbAgGe ($H \| ab$), with nuclear
contributions subtracted,  plotted as $\Delta C_p/T$ {\it vs.} $T$ on a semi-log plot. Arrow marks magnetic
transition in $H = 4$ T curve; (b) Low temperature differential susceptibility, $dM/dH$, and Sommerfeld
coefficient, $C/T$, with nuclear contributions subtracted, as a function of applied magnetic field. Data for $T =
0.25$ K are shifted vertically by 0.1 $\mu_B$/T ($dM/dH$) and 0.825 J/mol K$^2$ ($C/T$) for clarity.}\label{fi7}
\end{figure}

From the $C_p(T)$ data taken in different fields we can assemble $C(H)/T$ data and plot them together with the
differential susceptibility, $\chi = dM/dH$. Fig. \ref{fi4} illustrates that, at least over some $H - T$ range,
these two quantities are proportional to each other, with a constant proportionality factor. This implies that the
$C/T$ data as well as the $dM/dH$ data emerge from the same electronic degrees of freedom. It should be noted that
the magnetic transition is at least as clearly seen in the $C/T$ {\it vs.} $H$ as in the $dM/dH$ curves (Fig.
\ref{fi4}). In order to follow this proportionality to lower temperatures the nuclear contributions to the
specific heat (Schottky, hyperfine) were subtracted. The nuclear specific heat $C_{nuc}=\alpha /T^2$ was
determined by fitting $CT^2$ data at low temperatures, below ~0.1\ K, with $\alpha+\gamma T^3$. The resulting data
are shown in Figure \ref{fi7}. As can be seen, below 300 mK the $\Delta C_p/T$ data are essentially
temperature-independent (Fig. \ref{fi7}(a), inset). It should be noted that within the framework of this
background subtraction, this temperature independence of $\Delta C_p/T$ is in contrast to the clear linear in
temperature resistivity \cite{nik05a} in parts of this $H - T$ region.  This allows for the estimation of $\Delta
C_p/T$ for $T = 0.05$ K. These data are compared to $dM/dH$ at the same temperature. As can be seen in Fig.
\ref{fi7}(b), the proportionality between $C_p/T$ and $dM/dH$ appears to extend down to our lowest measured
temperatures. On the other hand, it breaks down for temperatures of 1 K and higher (Fig. \ref{fi4}). This is the
same temperature range in which the temperature-dependent resistivity measured in different applied magnetic
fields allows for a single-power-law fit \cite{nik05a}. It should be noted that an accurate assessment of the
Sommerfeld - Wilson ratio, $R = 4 \chi \pi^2 k_B^2/3 \gamma \mu_{eff}^2$, from the data in Fig. \ref{fi4} is
hampered by the anisotropy of the magnetic susceptibility in YbAgGe and ambiguity of the $\mu_{eff}$ value in the
$H - T$ domain of the aforementioned data. If forced to do it, the evaluation using the high temperature
$\mu_{eff} = 4.54 \mu_B$ (an overestimate for the low temperature $\mu_{eff}$) will result in the (underestimated)
Sommerfeld - Wilson ratio value $R \approx 1$.

Further examination of the Fig. \ref{fi4} reveals the shoulder at $\sim 6 - 7$ T in the $C/T$ data (as well as in
the base temperature, 50 mK, $dM/dH$ data in Fig. \ref{fi2}(b)). A similar anomaly in $dM/dH$ and $\gamma(H)$,
observed in YbRh$_2$Si$_2$ \cite{tok05a}, was associated with a suppression of the heavy fermion state; this was
coincident with the saturation of the magnetization and it was argued that this helped confirm the  field-induced
localization of $f$-electrons. Such an association is not appropriate for YbAgGe where the saturation of the
magnetization data is not yet reached by 11 T (Fig. \ref{fi2}(a)). The shoulder in Fig. \ref{fi4} may be a sign of
some field-induced Fermi surface changes. Additionally, it is worth mentioning that the Sommerfeld - Wilson ratio
is usually constant when the sample is in the Fermi-liquid regime. Earlier thermodynamic and transport
measurements in YbAgGe \cite{bud04a,nik05a,bud05b} suggest that for $H \| ab$ the Fermi-liquid regime in this
material sets up at $H = 9 - 10$ T, significantly above the field of 4.5 - 5 T at and above which the constant
Sommerfeld - Wilson ratio is observed (Fig. \ref{fi4}). Additional theoretical and experimental efforts are
required for further understanding of the ground state of YbAgGe in the intermediate field range, but it appears
that the features in $dM/dH$ as well as $C/T$ that occur in the non-Fermi liquid and Fermi-liquid regimes
originate from the same electronic degrees of freedom.

Fig. \ref{fi3} shows  the $H - T$ phase diagram for YbAgGe ($H \| ab$) \cite{bud05b,nik05a} overlaid by the
aforementioned features. All magnetic phase lines are seen in the magnetization measurements. The hysteresis of
the lower magnetic transition increases with decrease of temperature and magnetic field. The Hall line that
originates at the field-induced QCP \cite{bud05a,bud05b} appears to have a corresponding, broad feature in
$dM/dH$, that very poorly manifests itself in the raw $M(H)$ data. The Hall line at higher fields coincides with
the shoulders in both the $dM/dH$ and the $C/T$ data. At this point these phase lines consistently manifest
themselves in magnetization, specific heat, magnetoresistance and Hall effect measurements.

\section{Summary}

Temperature- and field-dependent magnetization measurements performed on YbAgGe single crystals with the magnetic
field in the basal plane confirm the magnetic phase lines suggested by the electrical transport and specific heat
measurements in applied field. The Hall line that originates at the field-induced QCP \cite{bud05a,bud05b} has a
correspondent feature in $dM/dH$ data as well, and this feature extends to temperatures above the zero applied
field  magnetic ordering temperature. For all studied temperature range (50 mK $\leq T \leq$ 2 K) the
magnetization does not saturate below 11 T, suggesting that the Yb moment is not saturated in the $H - T$ domain
of these measurements, and the Yb magnetic moment saturation is not associated with the field induced QCP in this
material. The observed proportionality of the differential susceptibility and the linear-in-temperature
coefficient of specific heat clearly indicates that the same electronic degrees of freedom are responsible for the
features seen in both sets of data not only in the Fermi liquid regime, but also in the lower field, non-Fermi
liquid region.

\begin{acknowledgments}
Ames Laboratory is operated for the U.S. Department of Energy by Iowa State University under Contract No.
W-7405-Eng.-82. Work at Ames Laboratory was supported by the Director for Energy Research, Office of Basic Energy
Sciences. A. Pikul is indebted to the Alexander von Humbold Foundation for a research fellowship.
\end{acknowledgments}

\clearpage

\clearpage

\clearpage

\begin{figure}
\begin{center}
\includegraphics[angle=0,width=120mm]{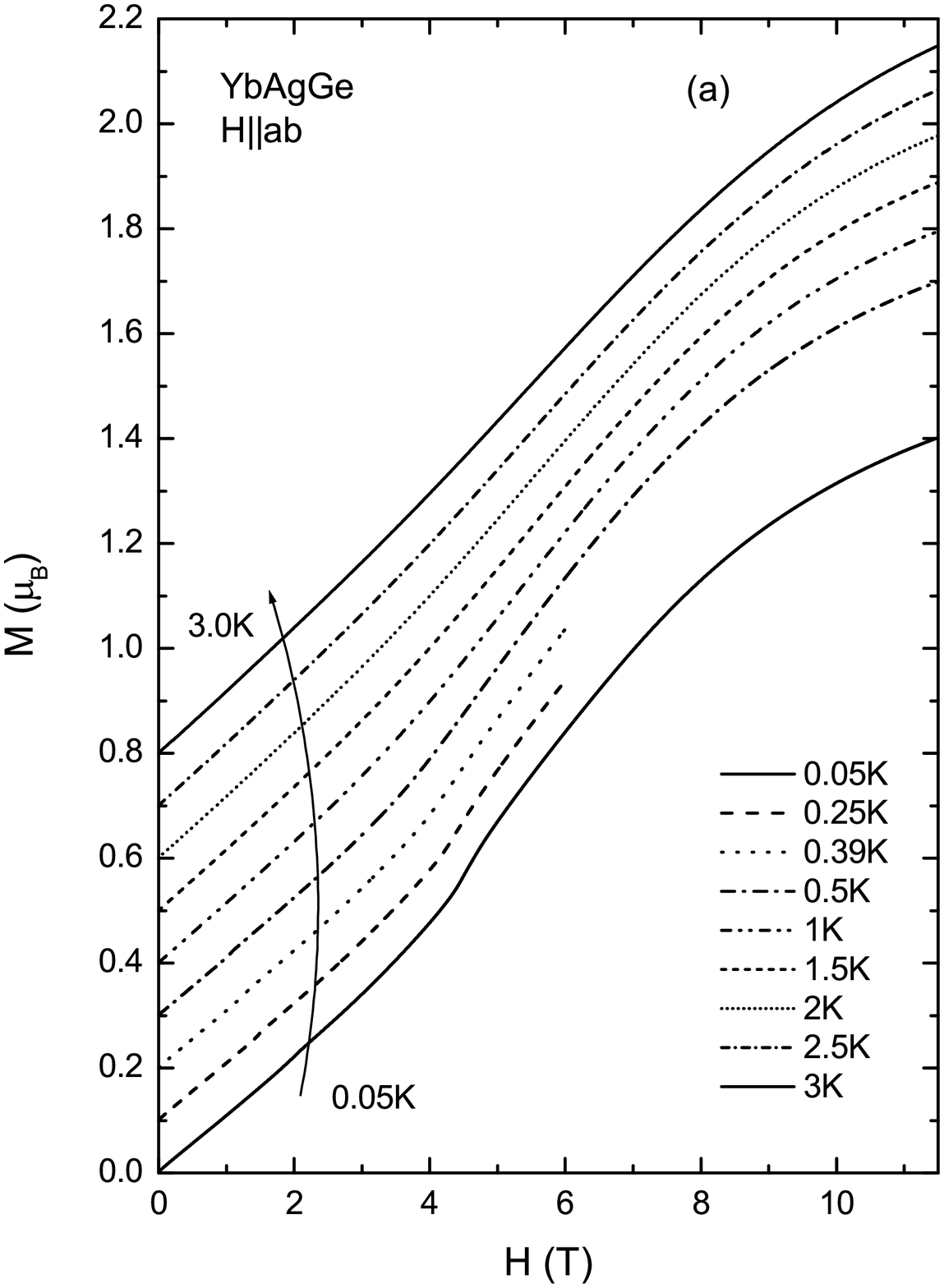}
\end{center}
\end{figure}

\clearpage

\clearpage

\clearpage

\clearpage

\clearpage

\begin{figure}
\begin{center}
\includegraphics[angle=0,width=120mm]{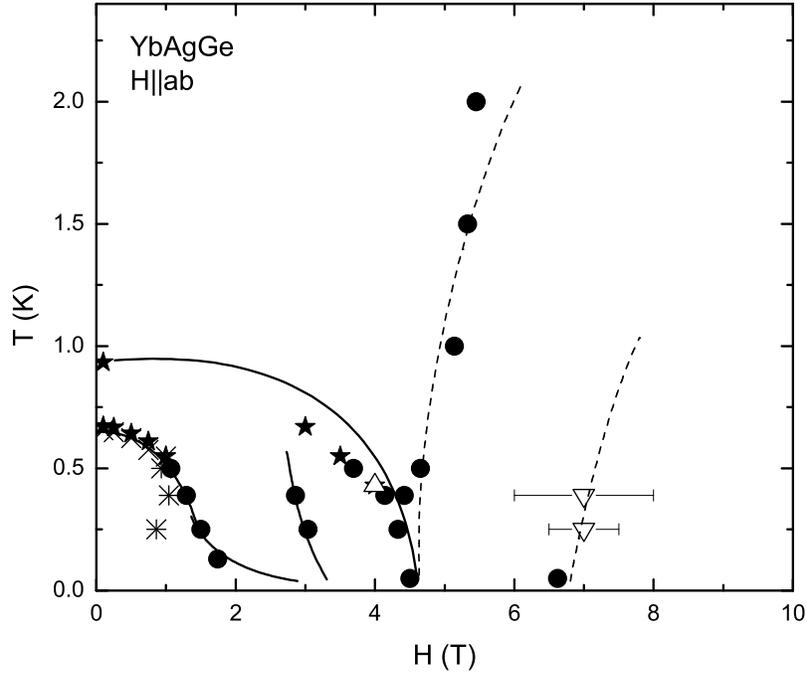}
\end{center}
\caption{$H - T$ phase diagram of YbAgGe ($H \| ab$): solid lines: magnetic phase lines \cite{bud04a,nik05a},
dashed lines - Hall effect lines \cite{bud05a,bud05b}, symbols - data from this work: $\bullet$ - from field-up
$M(H)$ data, $\star$ - from $M(T)$ data taken on increase of temperature, $\ast$ - from field-down $M(H)$ data,
$\times$ - from $M(T)$ data taken on temperature decrease, and $\triangle$, $\triangledown$ - from heat capacity
data plotted as $C_p(T)$ and $C(H)/T$ respectively.}\label{fi3}
\end{figure}


\begin{thebibliography}{16}
\expandafter\ifx\csname natexlab\endcsname\relax\def\natexlab#1{#1}\fi \expandafter\ifx\csname
bibnamefont\endcsname\relax
  \def\bibnamefont#1{#1}\fi
\expandafter\ifx\csname bibfnamefont\endcsname\relax
  \def\bibfnamefont#1{#1}\fi
\expandafter\ifx\csname citenamefont\endcsname\relax
  \def\citenamefont#1{#1}\fi
\expandafter\ifx\csname url\endcsname\relax
  \def\url#1{\texttt{#1}}\fi
\expandafter\ifx\csname urlprefix\endcsname\relax\def\urlprefix{URL }\fi \providecommand{\bibinfo}[2]{#2}
\providecommand{\eprint}[2][]{\url{#2}}

\bibitem[{\citenamefont{Bud'ko et~al.}(2004)\citenamefont{Bud'ko, Morosan, and
  Canfield}}]{bud04a}
\bibinfo{author}{\bibfnamefont{S.~L.} \bibnamefont{Bud'ko}},
  \bibinfo{author}{\bibfnamefont{E.}~\bibnamefont{Morosan}}, \bibnamefont{and}
  \bibinfo{author}{\bibfnamefont{P.~C.} \bibnamefont{Canfield}},
  \bibinfo{journal}{Phys.\ Rev.\ B} \textbf{\bibinfo{volume}{69}},
  \bibinfo{pages}{014415} (\bibinfo{year}{2004}).

\bibitem[{\citenamefont{Morosan et~al.}(2004)\citenamefont{Morosan, Bud'ko,
  Canfield, Torikachvili, and Lacerda}}]{mor04a}
\bibinfo{author}{\bibfnamefont{E.}~\bibnamefont{Morosan}},
  \bibinfo{author}{\bibfnamefont{S.~L.} \bibnamefont{Bud'ko}},
  \bibinfo{author}{\bibfnamefont{P.~C.} \bibnamefont{Canfield}},
  \bibinfo{author}{\bibfnamefont{M.~S.} \bibnamefont{Torikachvili}},
  \bibnamefont{and} \bibinfo{author}{\bibfnamefont{A.~H.}
  \bibnamefont{Lacerda}}, \bibinfo{journal}{J.\ Magn.\ Magn.\ Mat.}
  \textbf{\bibinfo{volume}{277}}, \bibinfo{pages}{298} (\bibinfo{year}{2004}).

\bibitem[{\citenamefont{Umeo et~al.}(2004)\citenamefont{Umeo, Yamane, Muro,
  Katoh, Niide, Ochiai, Morie, Sakakibara, and Takabatake}}]{ume04a}
\bibinfo{author}{\bibfnamefont{K.}~\bibnamefont{Umeo}},
  \bibinfo{author}{\bibfnamefont{K.}~\bibnamefont{Yamane}},
  \bibinfo{author}{\bibfnamefont{Y.}~\bibnamefont{Muro}},
  \bibinfo{author}{\bibfnamefont{K.}~\bibnamefont{Katoh}},
  \bibinfo{author}{\bibfnamefont{Y.}~\bibnamefont{Niide}},
  \bibinfo{author}{\bibfnamefont{A.}~\bibnamefont{Ochiai}},
  \bibinfo{author}{\bibfnamefont{T.}~\bibnamefont{Morie}},
  \bibinfo{author}{\bibfnamefont{T.}~\bibnamefont{Sakakibara}},
  \bibnamefont{and}
  \bibinfo{author}{\bibfnamefont{T.}~\bibnamefont{Takabatake}},
  \bibinfo{journal}{J.\ Phys.\ Soc.\ Jpn.} \textbf{\bibinfo{volume}{73}},
  \bibinfo{pages}{537} (\bibinfo{year}{2004}).

\bibitem[{\citenamefont{Katoh et~al.}(2004)\citenamefont{Katoh, Mano, Nakano,
  Terui, Niide, and Ochiai}}]{kat04a}
\bibinfo{author}{\bibfnamefont{K.}~\bibnamefont{Katoh}},
  \bibinfo{author}{\bibfnamefont{Y.}~\bibnamefont{Mano}},
  \bibinfo{author}{\bibfnamefont{K.}~\bibnamefont{Nakano}},
  \bibinfo{author}{\bibfnamefont{G.}~\bibnamefont{Terui}},
  \bibinfo{author}{\bibfnamefont{Y.}~\bibnamefont{Niide}}, \bibnamefont{and}
  \bibinfo{author}{\bibfnamefont{A.}~\bibnamefont{Ochiai}},
  \bibinfo{journal}{J.\ Magn.\ Magn.\ Mat.} \textbf{\bibinfo{volume}{268}},
  \bibinfo{pages}{212} (\bibinfo{year}{2004}).

\bibitem[{\citenamefont{Matsumura et~al.}(2004)\citenamefont{Matsumura, Ishida,
  Sato, Katoh, Niide, and Ochiai}}]{mat04a}
\bibinfo{author}{\bibfnamefont{T.}~\bibnamefont{Matsumura}},
  \bibinfo{author}{\bibfnamefont{H.}~\bibnamefont{Ishida}},
  \bibinfo{author}{\bibfnamefont{T.~J.} \bibnamefont{Sato}},
  \bibinfo{author}{\bibfnamefont{K.}~\bibnamefont{Katoh}},
  \bibinfo{author}{\bibfnamefont{Y.}~\bibnamefont{Niide}}, \bibnamefont{and}
  \bibinfo{author}{\bibfnamefont{A.}~\bibnamefont{Ochiai}},
  \bibinfo{journal}{J.\ Phys.\ Soc.\ Jpn.} \textbf{\bibinfo{volume}{73}},
  \bibinfo{pages}{2967} (\bibinfo{year}{2004}).

\bibitem[{\citenamefont{Bud'ko et~al.}(2005)\citenamefont{Bud'ko, Morosan, and
  Canfield}}]{bud05a}
\bibinfo{author}{\bibfnamefont{S.~L.} \bibnamefont{Bud'ko}},
  \bibinfo{author}{\bibfnamefont{E.}~\bibnamefont{Morosan}}, \bibnamefont{and}
  \bibinfo{author}{\bibfnamefont{P.~C.} \bibnamefont{Canfield}},
  \bibinfo{journal}{Phys.\ Rev.\ B} \textbf{\bibinfo{volume}{71}},
  \bibinfo{pages}{054408} (\bibinfo{year}{2005}).

\bibitem[{\citenamefont{Niklowitz et~al.}()\citenamefont{Niklowitz, Knebel,
  Flouquet, Bud'ko, and Canfield}}]{nik05a}
\bibinfo{author}{\bibfnamefont{P.~G.} \bibnamefont{Niklowitz}},
  \bibinfo{author}{\bibfnamefont{G.}~\bibnamefont{Knebel}},
  \bibinfo{author}{\bibfnamefont{J.}~\bibnamefont{Flouquet}},
  \bibinfo{author}{\bibfnamefont{S.~L.} \bibnamefont{Bud'ko}},
  \bibnamefont{and} \bibinfo{author}{\bibfnamefont{P.~C.}
  \bibnamefont{Canfield}}, \bibinfo{note}{cond-mat/0507211}.

\bibitem[{\citenamefont{Bud'ko et~al.}()\citenamefont{Bud'ko, Zapf, Morosan,
  and Canfield}}]{bud05b}
\bibinfo{author}{\bibfnamefont{S.~L.} \bibnamefont{Bud'ko}},
  \bibinfo{author}{\bibfnamefont{V.}~\bibnamefont{Zapf}},
  \bibinfo{author}{\bibfnamefont{E.}~\bibnamefont{Morosan}}, \bibnamefont{and}
  \bibinfo{author}{\bibfnamefont{P.~C.} \bibnamefont{Canfield}},
  \bibinfo{note}{cond-mat/0507338}.

\bibitem[{\citenamefont{F{\aa}k et~al.}(2005)\citenamefont{F{\aa}k, McMorrow,
  Niklowitz, Raymond, Ressouche, Flouquet, Canfield, Bud'ko, Janssen, and
  Gutmann}}]{fak05a}
\bibinfo{author}{\bibfnamefont{B.}~\bibnamefont{F{\aa}k}},
  \bibinfo{author}{\bibfnamefont{D.~F.} \bibnamefont{McMorrow}},
  \bibinfo{author}{\bibfnamefont{P.~G.} \bibnamefont{Niklowitz}},
  \bibinfo{author}{\bibfnamefont{S.}~\bibnamefont{Raymond}},
  \bibinfo{author}{\bibfnamefont{E.}~\bibnamefont{Ressouche}},
  \bibinfo{author}{\bibfnamefont{J.}~\bibnamefont{Flouquet}},
  \bibinfo{author}{\bibfnamefont{P.~C.} \bibnamefont{Canfield}},
  \bibinfo{author}{\bibfnamefont{S.~L.} \bibnamefont{Bud'ko}},
  \bibinfo{author}{\bibfnamefont{Y.}~\bibnamefont{Janssen}}, \bibnamefont{and}
  \bibinfo{author}{\bibfnamefont{M.~J.} \bibnamefont{Gutmann}},
  \bibinfo{journal}{J.\ Phys.:\ Condens.\ Matter}
  \textbf{\bibinfo{volume}{17}}, \bibinfo{pages}{301} (\bibinfo{year}{2005}).

\bibitem[{\citenamefont{F{\aa}k et~al.}()\citenamefont{F{\aa}k, Niklowitz,
  McMorrow, R{\"u}egg, Raymond, Flouquet, Canfield, Bud'ko, and
  Janssen}}]{fak05b}
\bibinfo{author}{\bibfnamefont{B.}~\bibnamefont{F{\aa}k}},
  \bibinfo{author}{\bibfnamefont{P.~G.} \bibnamefont{Niklowitz}},
  \bibinfo{author}{\bibfnamefont{D.~F.} \bibnamefont{McMorrow}},
  \bibinfo{author}{\bibfnamefont{C.}~\bibnamefont{R{\"u}egg}},
  \bibinfo{author}{\bibfnamefont{S.}~\bibnamefont{Raymond}},
  \bibinfo{author}{\bibfnamefont{J.}~\bibnamefont{Flouquet}},
  \bibinfo{author}{\bibfnamefont{P.~C.} \bibnamefont{Canfield}},
  \bibinfo{author}{\bibfnamefont{S.~L.} \bibnamefont{Bud'ko}},
  \bibnamefont{and} \bibinfo{author}{\bibfnamefont{Y.}~\bibnamefont{Janssen}},
  \bibinfo{note}{in: Proceedings of SCES'05, Vienna}.

\bibitem[{\citenamefont{Continentino et~al.}(1989)\citenamefont{Continentino,
  Japiassu, and Troper}}]{con89a}
\bibinfo{author}{\bibfnamefont{M.~A.} \bibnamefont{Continentino}},
  \bibinfo{author}{\bibfnamefont{G.}~\bibnamefont{Japiassu}}, \bibnamefont{and}
  \bibinfo{author}{\bibfnamefont{A.}~\bibnamefont{Troper}},
  \bibinfo{journal}{Phys.\ Rev.\ B} \textbf{\bibinfo{volume}{39}},
  \bibinfo{pages}{9734} (\bibinfo{year}{1989}).

\bibitem[{\citenamefont{Paschen et~al.}(2004)\citenamefont{Paschen,
  L{\"u}hmann, Wirth, Gegenwart, Trovarelli, Geibel, Steglich, Coleman, and
  Si}}]{pas04a}
\bibinfo{author}{\bibfnamefont{S.}~\bibnamefont{Paschen}},
  \bibinfo{author}{\bibfnamefont{T.}~\bibnamefont{L{\"u}hmann}},
  \bibinfo{author}{\bibfnamefont{S.}~\bibnamefont{Wirth}},
  \bibinfo{author}{\bibfnamefont{P.}~\bibnamefont{Gegenwart}},
  \bibinfo{author}{\bibfnamefont{O.}~\bibnamefont{Trovarelli}},
  \bibinfo{author}{\bibfnamefont{C.}~\bibnamefont{Geibel}},
  \bibinfo{author}{\bibfnamefont{F.}~\bibnamefont{Steglich}},
  \bibinfo{author}{\bibfnamefont{P.}~\bibnamefont{Coleman}}, \bibnamefont{and}
  \bibinfo{author}{\bibfnamefont{Q.}~\bibnamefont{Si}},
  \bibinfo{journal}{Nature} \textbf{\bibinfo{volume}{432}},
  \bibinfo{pages}{881} (\bibinfo{year}{2004}).

\bibitem[{\citenamefont{Mozharivskyj}()}]{moz04a}
\bibinfo{author}{\bibfnamefont{Y.}~\bibnamefont{Mozharivskyj}},
  \bibinfo{note}{private communication}.

\bibitem[{\citenamefont{Sakakibara et~al.}(1994)\citenamefont{Sakakibara,
  Mitamura, Tayama, and Amitsuka}}]{sak94a}
\bibinfo{author}{\bibfnamefont{T.}~\bibnamefont{Sakakibara}},
  \bibinfo{author}{\bibfnamefont{H.}~\bibnamefont{Mitamura}},
  \bibinfo{author}{\bibfnamefont{T.}~\bibnamefont{Tayama}}, \bibnamefont{and}
  \bibinfo{author}{\bibfnamefont{H.}~\bibnamefont{Amitsuka}},
  \bibinfo{journal}{Jpn.\ J.\ Appl.\ Phys.} \textbf{\bibinfo{volume}{33}},
  \bibinfo{pages}{5067} (\bibinfo{year}{1994}).

\bibitem[{\citenamefont{Wilhelm et~al.}(2004)\citenamefont{Wilhelm,
  L{\"u}hmann, Rus, and Steglich}}]{wil04a}
\bibinfo{author}{\bibfnamefont{H.}~\bibnamefont{Wilhelm}},
  \bibinfo{author}{\bibfnamefont{T.}~\bibnamefont{L{\"u}hmann}},
  \bibinfo{author}{\bibfnamefont{T.}~\bibnamefont{Rus}}, \bibnamefont{and}
  \bibinfo{author}{\bibfnamefont{F.}~\bibnamefont{Steglich}},
  \bibinfo{journal}{Rev.\ Sci.\ Instr.} \textbf{\bibinfo{volume}{75}},
  \bibinfo{pages}{2700} (\bibinfo{year}{2004}).

\bibitem[{\citenamefont{Tokiwa et~al.}(2005)\citenamefont{Tokiwa, Gegenwart,
  Radu, Ferstl, Sparn, Geibel, and Steglich}}]{tok05a}
\bibinfo{author}{\bibfnamefont{Y.}~\bibnamefont{Tokiwa}},
  \bibinfo{author}{\bibfnamefont{P.}~\bibnamefont{Gegenwart}},
  \bibinfo{author}{\bibfnamefont{T.}~\bibnamefont{Radu}},
  \bibinfo{author}{\bibfnamefont{J.}~\bibnamefont{Ferstl}},
  \bibinfo{author}{\bibfnamefont{G.}~\bibnamefont{Sparn}},
  \bibinfo{author}{\bibfnamefont{C.}~\bibnamefont{Geibel}}, \bibnamefont{and}
  \bibinfo{author}{\bibfnamefont{F.}~\bibnamefont{Steglich}},
  \bibinfo{journal}{Phys.\ Rev.\ Lett.} \textbf{\bibinfo{volume}{94}},
  \bibinfo{pages}{226402} (\bibinfo{year}{2005}).

\end{thebibliography}
\end{document}